\documentclass[a4paper]{jpconf}

\usepackage{graphicx}
\usepackage{times}
\usepackage{txfonts}
\usepackage{iopams}
\usepackage{amssymb}

\usepackage{color}
\usepackage{subfigure}
\usepackage{ulem}
\newcommand{\ep}{\varepsilon}

\newcommand{\eqs}[1]{\begin{equation}\begin{eqalign}   #1  \end{eqalign} \end{equation}}

\newcommand{\ks}[1]{#1 \!\!\! \slash } 
\newcommand{\kks}[1]{#1 \!\! \slash } 

\newcommand{\ga}{\gamma^5}

\newcommand{\ie}{{\it i.e.}}

\newcommand{\I}{{\cal I}}
\newcommand{\Ip}{{\cal I}'}

\newcommand{\ce}[1]{Eq.~(\ref{#1})}

\newcommand{\cf}[1]{{Fig.~\ref{#1}}}

\newcommand{\nn}{\nonumber}

\usepackage{ifpdf}
\ifpdf
\usepackage[pdftex]{hyperref}
\else
\usepackage[hypertex]{hyperref}
\fi

\hypersetup{
  pdftitle={Spin Observables in Transition-Distribution-Amplitude Studies},%
  pdfauthor={J.P. Lansberg},%
  pdfsubject={},%
  pdfkeywords={},%
  pdfstartview={},%
  bookmarksopen=true, breaklinks=true, debug=true, %
  colorlinks=true, linkcolor=blue, citecolor=blue, urlcolor=blue
}

\newcommand{\beq}[1]{
\begin{equation}\label{#1}}
\newcommand{\eeq}{\end{equation}}
\newcommand{\bea}[1]{
\begin{eqnarray}\label{#1}}
\newcommand{\eea}{\end{eqnarray}}

\newcommand{\eqsa}[1]{\begin{eqnarray}#1\end{eqnarray} }
\newcommand{\eq}[1]{\begin{equation}#1\end{equation} }
\newcommand{\out}{\raise-3pt\hbox{\scriptsize    out}}    

\begin{document}
\title{Spin Observables in Transition-Distribution-Amplitude Studies}

\author{J.P. Lansberg$^{1,2,}$\footnote[99]{Permanent address at IPNO}, B. Pire$^2$, L. Szymanowski$^3$}

\address{$^{1}$ IPNO, Universit\'e Paris-Sud 11, CNRS/IN2P3, F-91406,  Orsay, France \\
$^2$ Centre de Physique Th\'eorique, \'Ecole polytechnique, CNRS, F-91128, Palaiseau, France\\
$^3$ Soltan Institute for Nuclear Studies, Warsaw, Poland}

\ead{Jean-Philippe.Lansberg@in2p3.fr, Bernard.Pire@cpht.polytechnique.fr, \hspace*{0.9cm} Lech.Szymanowski@fuw.edu.pl}

\begin{abstract}
Exclusive hadronic reactions with a massive lepton pair ($\ell^+\ell^-$) in the final state will be measured with
PANDA at GSI-FAIR and with Compass at CERN, both in $p+\bar p \to \ell^+\ell^-+\pi$ and 
$\pi+N\to  N'+\ell^+\ell^-$. Similarly, electroproduction of a meson in the backward region will be studied at JLAB.
We discuss here how the spin
structure of the amplitude for such processes will enable us to disentangle various mechanisms. For instance, target-transverse-spin
asymmetries are specific of a partonic description, where the amplitude is factorised in terms of baryon
to meson or meson to baryon Transition Distribution Amplitudes (TDAs) as opposed to what is expected from baryon-exchange
contributions.
\end{abstract}

\vspace*{-0.2cm}

\section{Introduction}

In \cite{Pire:2005ax,Pire:2005mt}, we have introduced the  concept of 
Transition Distribution Amplitudes (TDAs) containing new information on the hadron structure.
These non-perturbative objects appear in the study of exclusive reactions in  a 
new scaling regime, i.e. involving a large --timelike or spacelike-- 
$Q^2$ photon and a baryonic exchange in the $t$-channel.
This extends the concept of Generalised Parton Distributions (GPDs), as already advocated
in the pioneering work of \cite{Frankfurt:1999fp}.

In particular, we have discussed~\cite{Lansberg:2007ec} the backward electroproduction of a pion,
 \eq{\gamma^\star(q) N(p_1)  \to N'(p_2) \pi(p_\pi),}
on a proton (or neutron) target, whose amplitude can be factorised
in  a hard part convoluted with TDAs (see \cf{fig:fact} (a)). Along the same lines, 
 the reaction,
\eq{N(p_1) \bar N (p_2) \to \gamma^\star(q) \pi(p_\pi),}
in the  near forward region was also studied in terms of TDAs~\cite{Pire:2004ie,Lansberg:2007se,Lutz:2009ff}
(see \cf{fig:fact} (b)).
The TDAs involved in the description of Deeply-Virtual Compton Scattering (DVCS) 
in  the backward kinematics 
$ \gamma^\star (q) N(p_1)  \to N'(p_2) \gamma(p_\gamma)$
and the reaction
$N(p_1) \bar N (p_2) \to \gamma^\star(q) \gamma(p_\gamma)$ 
in the near forward region were given in \cite{Lansberg:2006uh}.

\begin{figure}[htb!]
\centering{
\subfigure[\scriptsize $\pi^0$ electroproduction at small $u$]{\includegraphics[width=0.28\textwidth,clip=true]{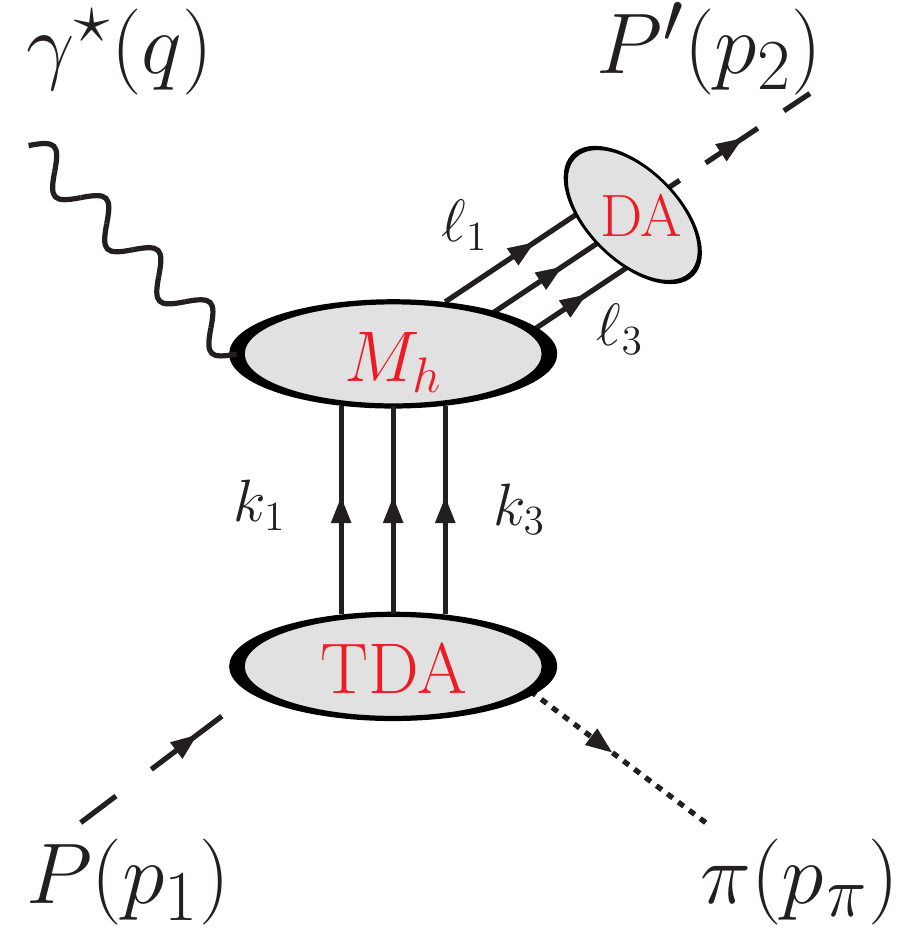}}
\subfigure[\scriptsize $p\bar p\to \gamma^\star \pi^0$ at small $t=(p_p-p_{\pi^0})^2$]{\includegraphics[width=0.28\textwidth,clip=true]{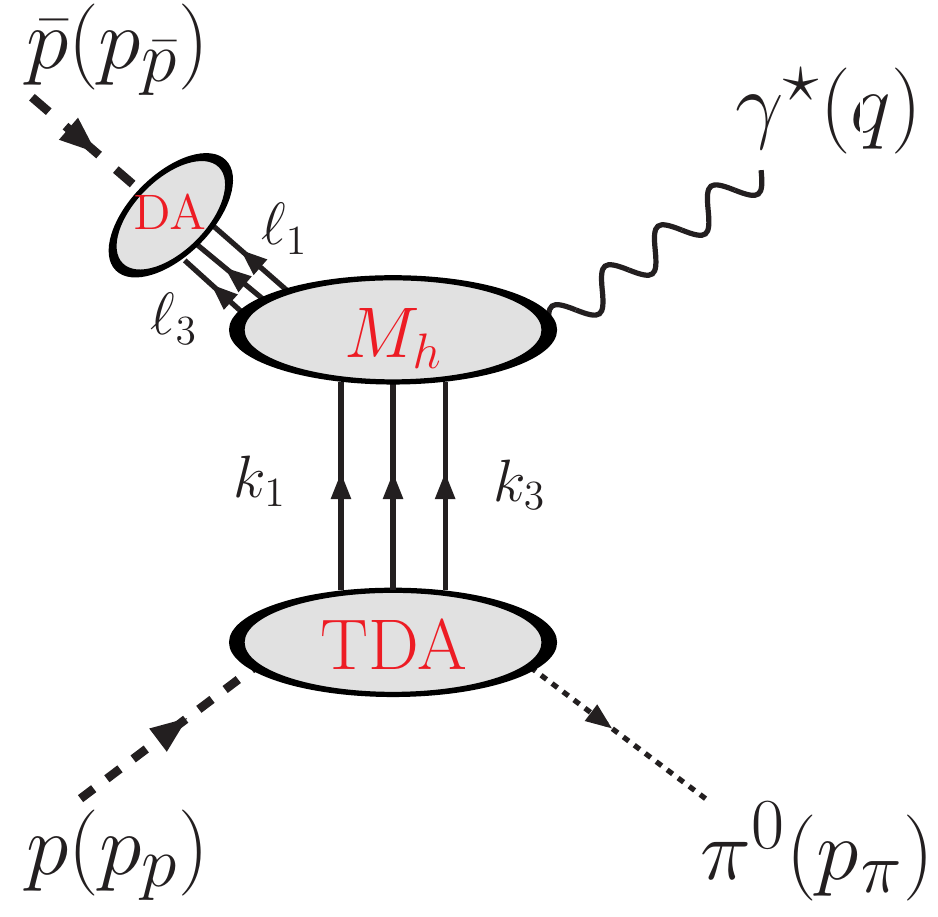}}
\subfigure[\scriptsize $p\bar p\to \gamma^\star \pi^0$ at small $u=(p_{\bar p}-p_{\pi^0})^2$]{\includegraphics[width=0.28\textwidth,clip=true]{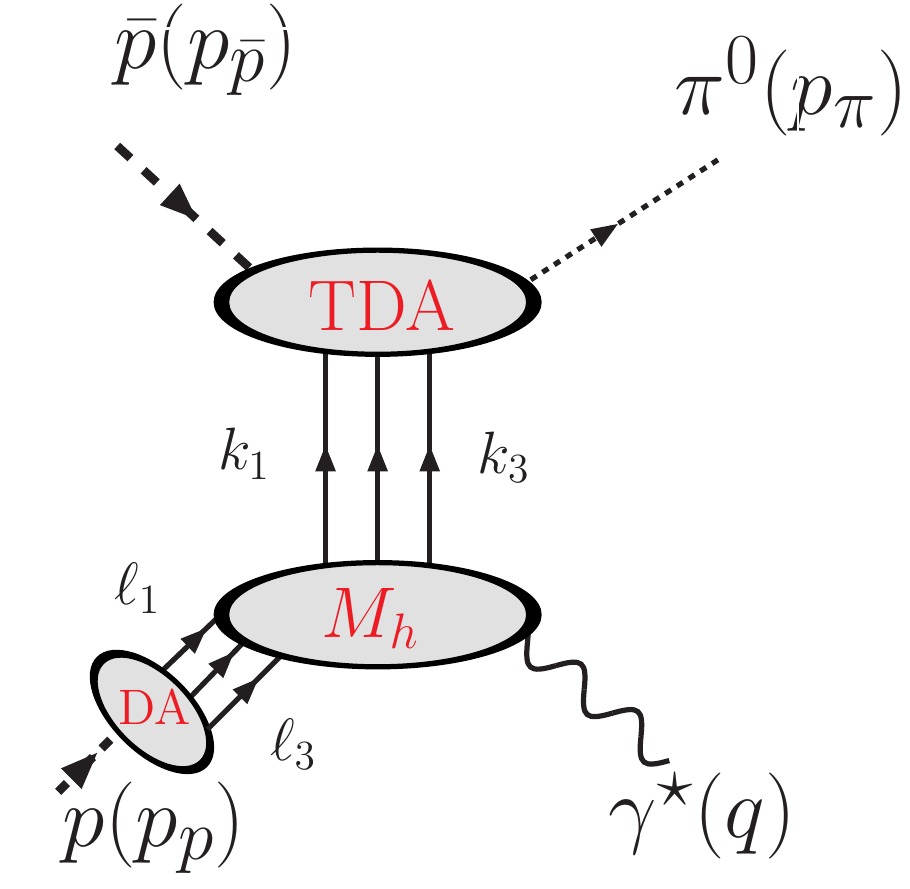}}
}
\caption{Illustration of the factorisation for three exclusive reactions involving the TDAs.}
\label{fig:fact}
\end{figure}

Recently, two studies of the proton-to-pion TDAs were carried out, one~\cite{Pasquini:2009ki} in the meson-cloud 
model, another~\cite{Pire:2010if} guided by the concept of the spectral 
representation~\cite{Radyushkin:1983wh,Radyushkin:1998es}. Yet, more work is needed
before being able to proceed to {\it quantitative} comparisons between different
TDA models and between theory and experiments. For the time being, model
independent analyses sound more expedient. These can be divided in three categories:
$Q^2$ dependence and scaling analyses, dominance of specific polarisations of the off-shell photon and
target (or projectile) transverse-spin asymmetries. In this note, we shall review these.

\section{$Q^2$-dependence, scaling and dominant photon polarisation: Example of the TDA studies in $\bar p p$ annihilation} 

Let us analyse here the case of $\bar p p \to \ell^+ \ell^- \pi$. This will enable us to discuss
the scaling and $Q^2$-dependence in the factorised picture as well as the dominance of specific photon polarisations.
The momenta of the subprocess $\bar p p \to \gamma^\star \pi$ are defined as shown in~\cf{fig:fact} (b). 
We shall first limit our discussion to the region where $t=(p_p-p_\pi)^2$ is much smaller than the invariant mass of the
$\ell^+ \ell^- $ pair, $Q^2$. The region where $u=(p_{\bar p}-p_\pi)^2\ll Q^2$ (see~\cf{fig:fact} (c)) is briefly discussed at the end of the section.

\subsection{Kinematics}
The $z$-axis is chosen along the colliding proton and anti-proton 
and the $x-z$ plane is identified 
with the collision or hadronic plane. We define the 
light-cone vectors $p$ and $n$ 
such that $2~p.n=1$, as well as $P= (p_p+p_\pi)/2$, $\Delta=p_\pi -p_p$ and its 
transverse component $\Delta_T$ ($\Delta_T^2<0$).  $\xi$ is defined as $\xi=-\frac{\Delta.n}{2P.n}$.
We express the particle momenta  through a 
 Sudakov decomposition, which,
for $\Delta_T=0$, $M\ll W$ and $m_\pi=0$, is
\eq{p_p=(1+\xi)p, ~~~p_{\bar p}=\frac{W^2}{1+\xi} n, ~~~p_\pi=(1-\xi)p, ~~~t=\frac{2\xi M^2}{1+\xi}, ~~~\xi=\frac{Q^2}{2W^2-Q^2}.}
In the fixed-target mode, the maximal reachable value for $W^2 = 2M^2 +2 M E_{\bar p}$ at GSI will 
be $\simeq 30$ GeV$^2$ (for $E_{\bar p}=15$ GeV). The highest invariant mass 
of the photon could be $Q^2_{max}\simeq 30$ GeV$^2$. We refer to~\cite{Adamuscin:2007iv}
for a complete discussion of the kinematically allowed domain. In terms of our notations, in the proton rest frame,
we have $p =\frac{M}{2(1+\xi)} (1,0,0,-1)$ and $n =\frac{1+\xi}{2M} (1,0,0,1)$. Thus 
 $\xi \in [0.5, 1]$ corresponds to $|p^z_{\pi}|<M/6 \simeq 155 $ MeV in the laboratory frame at 
$\Delta_T=0$.

\subsection{The properties of the amplitude}

At $\Delta_T=0$, the leading-twist TDAs for the $p \to \pi^0$ transition, $ V^{p\pi^0}_{i}\!\!(x_i,\xi, \Delta^2)$, 
$A^{p\pi^0}_{i}\!\!(x_i,\xi, \Delta^2)$ and 
$T^{p\pi^0}_{i}\!\!(x_i,\xi, \Delta^2)$  are defined  as:
\eqsa{\label{eq:TDApi0proton}
&  {\cal F}\Big(\langle     \pi^0(p_\pi)|\, \epsilon^{ijk}u^{i}_{\alpha}(z_1 n) 
u^{j}_{\beta}(z_2 n)d^{k}_{\gamma}(z_3 n)
\,|P(p_p,s_p) \rangle \Big)=   
\frac{i}{4}\frac{f_N}{f_\pi}\Big[ V^{p\pi^0}_{1} (\ks p C)_{\alpha\beta}(u^+(p_p,s_p))_{\gamma}& \nn \\
& +A^{p\pi^0}_{1} (\ks p\gamma^5 C)_{\alpha\beta}(\gamma^5 u^+(p_p,s_p))_{\gamma} 
 +T^{p\pi^0}_{1} (\sigma_{p\mu} C)_{\alpha\beta}(\gamma^\mu u^+(p_p,s_p))_{\gamma}\Big]\;,&\nn}
where  $\sigma^{\mu\nu}= 1/2[\gamma^\mu, \gamma^\nu]$, $C$ is the charge 
conjugation matrix, $f_\pi = 131$ MeV is the pion decay constant  and $f_N \sim 5.2\cdot 10^{-3}$ GeV$^2$. 
$u^+$ is the large component of the nucleon spinor such that one has  $u(p_p,s_p)=$ ~ $(\ks n \ks p + \ks p \ks n)u(p_p,s_p)  =  u^-(p_p,s_p)+u^+(p_p,s_p)$
with $u^+(p_p,s_p)\sim \sqrt{p_p^+}$ and $u^-(p_p,s_p)\sim \sqrt{1/p_p^+}$.
 
At the leading order in $\alpha_s$ and at $\Delta_T=0$, the amplitude 
${\cal M}_\lambda^{s_ps_{\bar p}}$  for 
$\bar p(p_{\bar p},s_{\bar p}) p(p_{p},s_p) \to \gamma^\star(q,\lambda) \pi^0(p_\pi)$ reads
\eq{\label{eq:ampl-ppbargammapi}
\!\!\!\!{\cal M}_\lambda^{s_p s_{\bar p}}=
-i 
\frac{(4 \pi \alpha_s)^2 \sqrt{4 \pi \alpha_{em}} f_{N}^2}{ 54 f_{\pi}Q^4} 
\underbrace{\bar v^+(p_{\bar p},s_{\bar p}) \ks \ep^\star(\lambda) \gamma^5 u^+(p_p,s_p)}_{{\cal S}_\lambda^{s_ps_{\bar p}}}
\underbrace{\int\limits^{1+\xi}_{-1+\xi} \!\!\!\! [dx]\!\! \int\limits_0^1 \!\![dy]
\Bigg(2\sum\limits_{\alpha=1}^{7}\! R_{\alpha}+\!\!\sum\limits_{\alpha=8}^{14}\! R_{\alpha}\Bigg)}_{\cal I},}
where $[dx]=dx_1 dx_2 dx_3\delta(2\xi - \sum_k x_k)$ and $[dy]=dy_1 dy_2 dy_3\delta(1 - \sum_k y_k)$; 
the coefficients $R_{\alpha}\,(\alpha=1,...,14)$ exactly correspond to $T_{\alpha}$
in~\cite{Lansberg:2007ec} after the replacement $-i\epsilon \to i \epsilon $ due to the presence 
of the $\gamma^\star$ in the final  instead of initial state.

Denoting the linear polarisations of the virtual photon  by the indices $L,x,y$,
one defines~\cite{Mulders:1990xw,Park:2007tn} $\sigma_T\propto 1/2 [{\cal M}_{{\tiny x}} ({\cal M}_{{\tiny x}})^\ast
+{\cal M}_{{\tiny y}} ({\cal M}_{{\tiny y}})^\ast]$ , $\sigma_L\propto {\cal M}_{{\tiny L}} ({\cal M}_{{\tiny L}})^\ast$,  
$\sigma_{TL}\propto {\cal M}_{{\tiny x}} ({\cal M}_{{\tiny L}})^\ast
+{\cal M}_{{\tiny L}} ({\cal M}_{{\tiny x}})^\ast$ and $\sigma_{TT}\propto
1/2 [{\cal M}_{{\tiny x}} ({\cal M}_{{\tiny x}})^\ast
-{\cal M}_{{\tiny y}} ({\cal M}_{{\tiny y}})^\ast]$. The corresponding definitions apply for the squared of
${{\cal S}_\lambda^{s_ps_{\bar p}}}$ summed over the proton spins, ${\cal S}_{T,L,LT,TT}^2$.

At the leading twist, only ${\cal S}_T^2= (2(1+\xi) Q^2) / \xi$ survives. This means that
the far off-shell photon produced in association with the pion is dominantly transversely 
polarised. This dominance increases with $Q^2$. As a direct consequence, the angular dependence of the lepton pair
follows the distribution \eq{1+\cos^2 \theta_\ell,} where $\theta_\ell$ is the angle between the $\ell^+$ momentum 
in the dilepton rest frame and the $z$ axis. Along the same line of argument, 
there should not be any azimuthal dependence, $\phi_\ell$, in the dilepton distribution. We should stress here that, at $|\Delta_T|^2=0$,
these results coincide with the ones for the proton form factor studies in the timelike region. However, for non-vanishing transverse momenta, 
the spin-quantisation axis may be rotated inducing  different definitions of the aforementioned angles.

Let us now discuss the scaling of the amplitude squared. While the contribution from ${\cal I}$ 
may depend on $\xi$ (and $\Delta_T^2$), it  does not depend on $Q^2$ (except for a logarithmic dependence of the TDAs if one takes into account their QCD evolution). 
Given that ${\cal S}_T^2= (2(1+\xi) Q^2) / \xi$, we deduce from \ce{eq:ampl-ppbargammapi} that the amplitude squared scales like $Q^{-6}$ at fixed $\xi$.

\subsection{The $\bar p \to \pi^0$ transition region}

So far, we have discussed the case where the momentum-transfer squared between the proton and the pion is small compared to $Q^2$, \ie~when the pion
is moving slowly in the proton-rest frame. However, 
nothing prevents us for applying the same arguments for the factorisation of the amplitude in 
the case where the pion is slow in the anti-proton rest frame. One expects the same kind of mechanism to take place. The amplitude
factorises into a hard part and anti-proton-to-pion TDAs. In practice, at GSI-FAIR, one would observe a very energetic (and thus near forward) 
pion and a very slow lepton pair in the laboratory frame. The same scaling properties
of the amplitude and the same expectations for the dilepton polarisation  would obviously hold.

\section{Single Transverse Spin Asymmetry: Example of the TDA studies in backward electroproduction of a pion }

In order to study a possible Single Spin Asymmetry (SSA) in backward electroproduction of a pion 
on a transversely polarised target,  we shall study the quantity $\sigma^{s_1}-\sigma^{-s_1}$  with
the definition
\eqs{\sigma^{s_1}=\sum_\lambda\sum_{s_2} ({\cal M}_\lambda^{s_1s_2})
({\cal M}_\lambda^{s_1s_2})^\ast.}
As we shall see the SSA can only arise for $\Delta_T\neq 0$ and will be a function of azimuthal angles of the process. 
We thus need to define precisely the kinematics  and say some words about  ${\cal M}_\lambda^{s_1s_2}$ for $\Delta_T\neq 0$.

\subsection{Kinematics}

The momenta of the process $\gamma^\star P \to P' \pi $ are defined as in \cf{fig:fact} (a) and \cf{fig:kin}.
The $z$-axis is chosen along the initial-nucleon and the virtual-photon momenta. The $x-z$ plane is identified 
with the collision or hadronic plane (\cf{fig:kin}). 
Here, $P=\frac{1}{2} (p_1+p_\pi)$, $\Delta=p_\pi -p_1$ and its 
transverse component is $\Delta_T$ ($\Delta_T.\Delta_T=\Delta_T^2<0$).

We  can then express the momenta of the particles through their  
Sudakov decomposition and, keeping the first-order corrections in the masses and $\Delta_T^2$, we have:
\eqsa{\label{eq:decomp_moment}
p_1&=&\! (1+\xi) p + \frac{M^2}{1+\xi}n, q\simeq\!- 2 \xi \Big(1+ \textstyle \frac{(\Delta_T^2-M^2)}{Q^2}\Big)  p + \frac{Q^2}{2\xi} \Big(1+ \textstyle \frac{(\Delta_T^2-M^2)}{Q^2}\Big)^{-1} n, \nn \\
p_\pi&=&\! (1-\xi) p +\frac{m_\pi^2-\Delta_T^2}{1-\xi}n+ \Delta_T,\Delta=\! - 2 \xi p +\Big[\frac{m_\pi^2-\Delta_T^2}{1-\xi}- \frac{M^2}{1+\xi}\Big]n
+ \Delta_T \nn\\
p_2&\simeq&- 2 \xi \frac{(\Delta_T^2-M^2)}{Q^2} p+\! \Big[\frac{Q^2}{2\xi} \Big(1+ \textstyle \frac{(\Delta_T^2-M^2)}{Q^2}\Big)^{-1} -\frac{m_\pi^2-\Delta_T^2}{1-\xi}+ \frac{M^2}{1+\xi}\Big]n - \Delta_T,
}
with $\xi\simeq\frac{Q^2}{Q^2+2(W^2+\Delta^2_T-M^2)}$.

For $\ep_x=(0,1,0,0)$ and $\ep_y=(0,0,1,0)$ with the axis definitions of \cf{fig:kin}, one may further 
specify that 
\eq{\Delta_T=|\Delta_T| (\cos \phi \, \ep_x + \sin \phi \, \ep_y) \hbox{ and }
s_{T,1}=s_{1}= \cos \phi_S \, \ep_x + \sin \phi_S \, \ep_y,}
for the transverse spin of the target ($s_1.p_1=s_1.p=s_1.n=0$).

$\theta^*_\pi$ is defined as the polar angle between the virtual photon and the pion in the $P'\pi^0$ center-of-mass frame (see \cf{fig:kin}).
$\phi$ is the azimuthal angle between the electron plane and the plane of the 
process $\gamma^\star P \to P' \pi^0$ (hadronic plane)
($\phi=0$ when the pion is emitted in the half plane containing the outgoing electron).

\begin{figure}[htb!]
\centering{
\includegraphics[width=0.8\textwidth,clip=true]{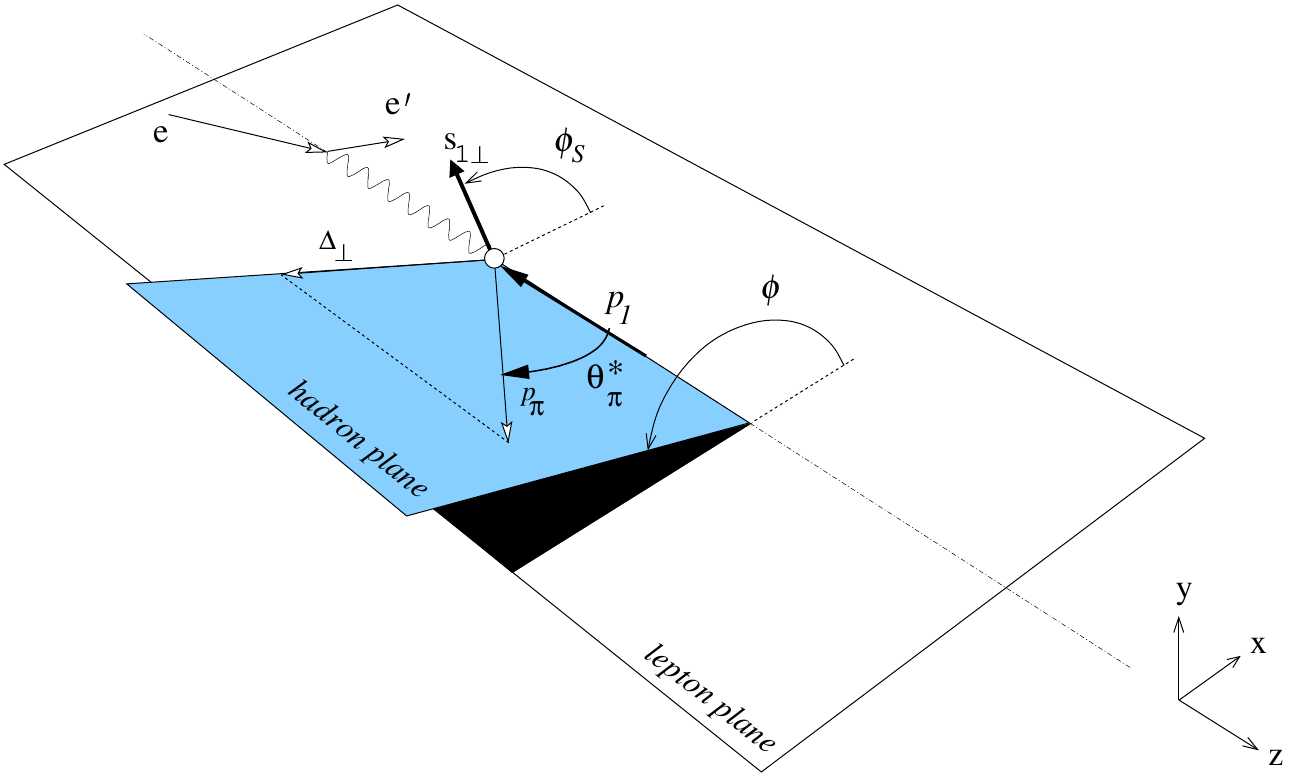}
} 
\caption{ Kinematics of electroproduction of a pion and definition of the angles $\phi$ and $\phi_S$}
\label{fig:kin}
\end{figure}

 \subsection{Hard-amplitude at $\Delta_T\neq 0$}

At leading order in $\alpha_s$, the amplitude ${\cal M}_\lambda^{s_1s_2}$  for 
$\gamma^\star(q,\lambda) P(p_1,s_1) \to P'(p_2,s_2) \pi^0(p_\pi)$ reads

\eqsa{\label{eq:ampl-bEPM1}
{\cal M}_\lambda^{s_1s_2}=\underbrace{-i 
\frac{(4 \pi \alpha_s)^2 \sqrt{4 \pi \alpha_{em}} f_{N}^2}{ 54 f_{\pi}}}_{\cal C}\frac{1}{Q^4} \!\!\!\!\!\!\!\!\!\!\!\! &\Big[ 
\underbrace{ \bar u_2 \ks \ep(\lambda) \gamma^5 u_1}_{{\cal S}_\lambda^{s_1s_2}}
\underbrace{\int \!
\Bigg(2\sum\limits_{\alpha=1}^{7} T_{\alpha}+
\sum\limits_{\alpha=8}^{14} T_{\alpha}\Bigg)}_{{\cal I}}\nn\\&-
\underbrace{ \bar u_2 \ks \ep(\lambda) \frac{\ks \Delta_{T}}{M} \gamma^5 u_1 }_{{\cal S'}_\lambda^{s_1s_2}}
\underbrace{\int \!
\Bigg(2\sum\limits_{\alpha=1}^{7} T'_{\alpha}+
\sum\limits_{\alpha=8}^{14} T'_{\alpha}\Bigg)}_{{\cal I'}}\Big]
,}
where $\int \equiv \int\limits^{1+\xi}_{-1+\xi} \! \! \! dx_1dx_2dx_3\delta(2\xi -x_1-x_2-x_3)
 \int\limits_0^1 \! \! dy_1dy_2dy_3\delta(1-y_1-y_2-y_3)$, 
$u(p_1,s_1)\equiv~u_1$,  $\bar u(p_2,s_2)\equiv\bar u_2$  and the coefficients $T_{\alpha}$ and $T'_{\alpha}$$(\alpha=1,...,14)$ are functions of 
$x_{i}$, $y_{j}$, $\xi$ and $\Delta^2$ and  are given in Table 1 
of \cite{Lansberg:2007ec}. 

The expression in \ce{eq:ampl-bEPM1} is to be compared with the leading-twist amplitude for the 
baryonic-form factor~\cite{CZ}
\eq{\label{eq:ampl-FF} {\cal M}_\lambda \propto  -i ( \bar u_2 \ks \ep(\lambda) u_1) \frac{\alpha_s^2 f_{N}^2}{Q^4}  \int
\Bigg(2\sum\limits_{\alpha=1}^{7} T^p_{\alpha}(x_{i},y_{j},\xi,t)+
\sum\limits_{\alpha=8}^{14} T^p_{\alpha}(x_{i},y_{j},\xi,t)\Bigg).}

The factors $T^p_{\alpha}$ are very similar to the $T_{\alpha}$ obtained here. However, the integration domain is different. 
In the form factor case
\eqs{\int \hbox{\ \ stands for \ \ } \int \limits_{0}^{1} dx_1dx_2dx_3\delta(1 -\sum_i x_i) \int 
\limits_0^1  dy_1dy_2dy_3\delta(1-\sum_i y_i).}
Consequently, the integration of denominators in $T^p_\alpha$ such as $1/(x_i+i \varepsilon)$ do not generate any
imaginary parts. On the contrary, the integrations of similar denominators in $T_\alpha$ and $T'_\alpha$ over the
 TDA integration domain will
generate an imaginary part when passing from the ERBL region (all $x_i>0$) to one of the DGLAP regions (one $x_i<0$).
This will be the source of the SSA as we will show later on. 

\subsection{The Single Transverse Spin Asymmetry}

Since we are interested in the leading twist contribution of this asymmetry, we can sum only over the transverse
polarisation of the virtual photon using 
${\sum}_{\lambda=x,y} \ep(\lambda)^\mu (\ep(\lambda)^\nu)^\ast=-g^{\mu\nu}+(p^\mu n^\nu+p^\nu n^\mu)/(p.n)$.
The sum on the final-proton spin $s_2$ is done using 
$ {\sum}_{s_2}u_\alpha(p_2,s_2) \bar u_\beta(p_2,s_2)=(\ks p_2+M)_{\alpha \beta}$. As regards the 
initial-proton spinor, one uses the following relation involving its  transverse spin $s_1$, 
$u_\alpha(p_1,s_1) \bar u_\beta(p_1,s_1)=1/2 (1+\ga \kks s_1)(\ks p_1+M)_{\alpha \beta}$.

Dropping the contributions proportional to the proton mass, the spin asymmetry reads 
\eqsa{\sigma^{s_1}-\sigma^{-s_1}&=& 8  \frac{|{\cal C}|^2}{Q^6} \frac{1+\xi}{\xi} \frac{\epsilon^{n p s_1 \Delta_T}}{M}  \Im m(\Ip \I^\ast)\\
&=&  - 4 \frac{|{\cal C}|^2}{Q^6} \frac{|\Delta_T|}{M} \frac{1+\xi}{\xi}  \sin(\phi-\phi_S) \Im m(\Ip \I^\ast).
}
Comparing with the expressions for the unpolarised cross section obtained in \cite{Lansberg:2007ec}, one concludes that
the asymmetry for the hard-parton induced contribution is leading-twist as soon as $\Delta_T\neq 0$ and $\I$ or $\Ip$ are no longer pure real or pure 
imaginary numbers. This is precisely what one expects when DGLAP contributions are taken into account~\cite{inprogress}.

\subsection{The Single Transverse Spin Asymmetry in $\bar p p$ annhilation}

Following the same reasonning, one can show that such an asymmetry would arise in $\bar p p \to \ell^+ \ell^- \pi^0$. Let us however emphasise
that the asymmetry on the (target) proton spin would then be leading twist only for the small-$t$ regime, while its projectile 
spin asymmetry is
leading twist in the small-$u$ regime. In the case where only the target can be polarised, SSA studies should then preferentially be carried out
by looking at slowly moving pions associated with a dilepton.

In general, spin asymmetries related to TDAs studies are of interest for the spin of the baryon which undergoes a non-perturbative transition
into a meson.

\begin{figure}[htb!]
\centering{
\subfigure[\scriptsize $p\bar p\to J/\psi \pi^0$ at small $t=(p_p-p_{\pi^0})^2$]{\includegraphics[width=0.3\textwidth,clip=true]{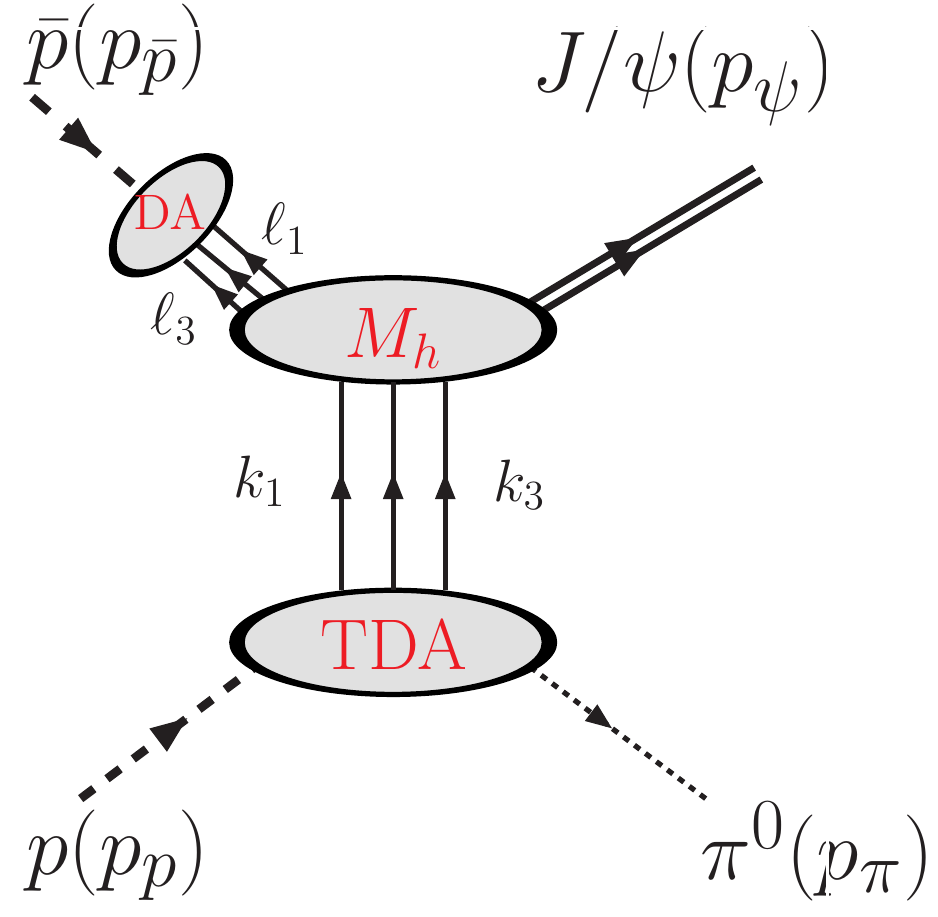}}
\subfigure[\scriptsize $\pi^- p\to \gamma^\star n$ at small $u=(p_{\pi}-p_n)^2$]{\includegraphics[width=0.3\textwidth,clip=true]{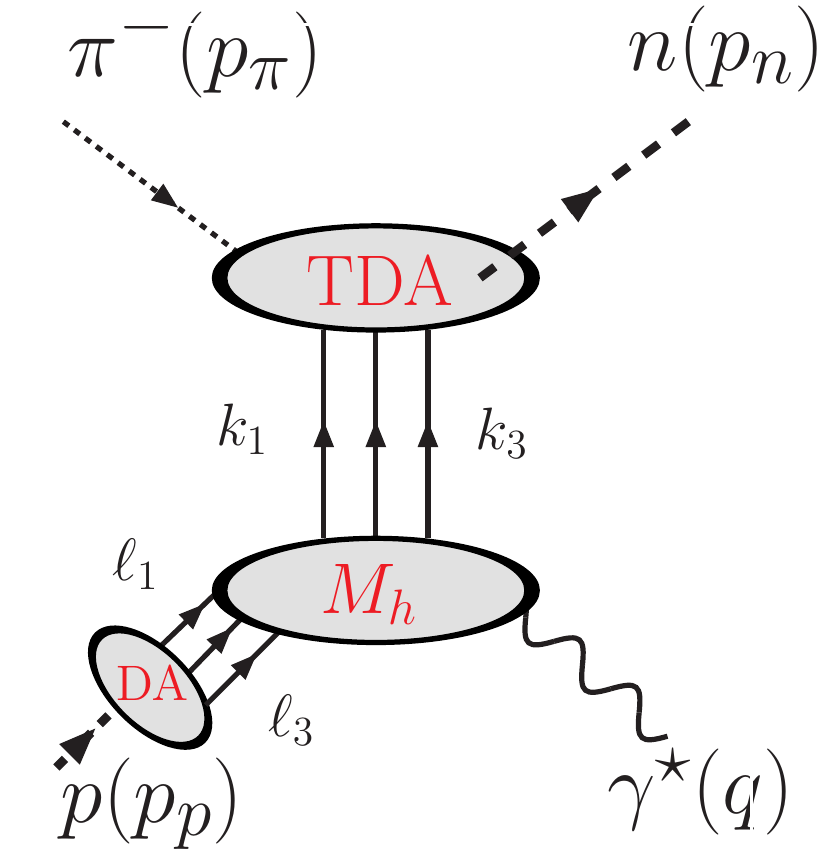}}
\subfigure[\scriptsize $K^- p\to \gamma^\star \Lambda$ at small $u=(p_{K}-p_\Lambda)^2$]{\includegraphics[width=0.3\textwidth,clip=true]{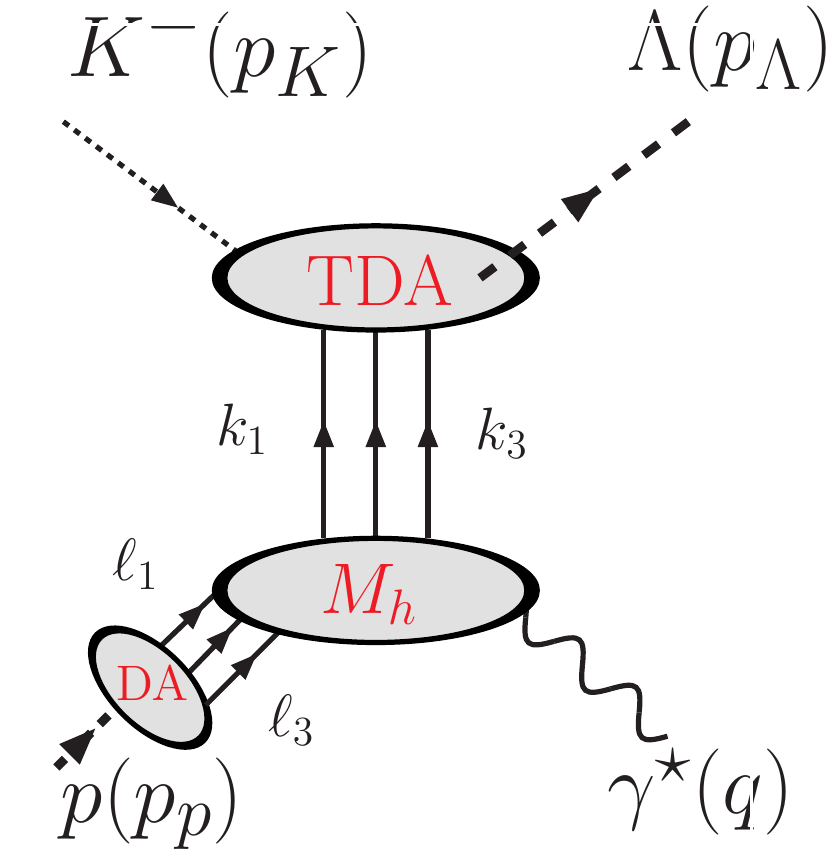}}
}
\caption{Three additional hard-exclusive reactions involving the TDAs at GSI-FAIR and COMPASS.}
\label{fig:fact2}
\end{figure}

\section{Other processes involving the TDAs}

\subsection{$\bar p p \to J/\psi \pi^0$ at GSI-FAIR} 
 
It is well-known that exclusive decay of $J/\psi$ into $p\bar p$ is rather well accounted for by the
pQCD mechanism involving DAs and a hard scattering with three gluons exchange between the $c \bar c$ pair
and the $(q \bar q)(q \bar q)(q \bar q)$ system~\cite{CZ,Chernyak:1987nv}. One then naturally expects  that
the production of $J/\psi$ in association with a pion could also be described
using the DAs and the TDAs on the one hand and the same hard scattering on the other. This is schematically
illustrated on \cf{fig:fact2} (a). Scaling properties and specific polarisation dominance should also appear
as well as SSA if the reaction is indeed occuring at the parton level and not at the level of the baryon and
the meson\cite{Gaillard:1982zm}.

\subsection{$\pi^- p \to n \gamma^\star$ and $K^- p \to \Lambda \gamma^\star$   at COMPASS}

Meson beams can also be used to study hard exclusive reaction in the timelike region. Due to
baryon-number conservation, one should still have a baryon in the final state. Using
a $\pi^-$, it was proposed~\cite{Berger:2001zn} to study $\pi^- p \to n \gamma^\star$ in the forward region --small 
$t=(p_{p}-p_n)^2$-- to extract information about the GPDs. Accordingly, it is perfectly reasonable
to think of the same study in the backward regime, where the final state baryon is very energetic
in the target-rest frame, more specifically at small $u=(p_{\pi}-p_n)^2$. For an outgoing photon
with sufficiently large $Q^2$, the amplitude should factorise in terms of pion-to-neutron TDAs
as depicted on \cf{fig:fact2} (b).

As for $\bar p p \to \gamma^\star \pi^0$, the outgoing photon would be transversely polarised.
However, the target transverse spin asymmetry would be higher twist since it is the neutron
which undergoes the non-perturbative transition to the $\pi^0$. Only the asymmetry of the
outgoing neutron spin would not be suppressed by any power of $Q^2$. 
To avoid this shortcoming, one could think of using $K^-$ beam. The corresponding process
would then be  $K^- p \to \Lambda \gamma^\star$, with a fast $\Lambda$ in the final state (see~\cf{fig:fact2} (c)).
Spin asymmetry on the latter could be then analysed through the azimuthal angular
dependence of the decay $\Lambda \to p \pi^-$.

\section{Discussion and conclusion}

Although the knowledge of {\it baryon to meson} TDAs has recently improved significantly thanks to 
a first study in the meson cloud model~\cite{Pasquini:2009ki} and another one focused on their spectral 
representations\cite{Pire:2010if}, model-independent observables aimed at studying the backward regime of
meson electroproduction or exclusive $\bar p p$ annhiliation into a meson and a dilepton 
will still be the bread-and-butter of this field for the months to come. 

Two obvious model-independent observables are the fixed $Q^{-6}$ dependence of the amplitude squared, for which the transverse polarisation
of the $\gamma^\star$ dominates, as well as the scaling in $\xi$, to be compared to $x_{Bj}$ in DIS.
We also find it particularly relevant to emphasise that the study of the asymmetry of the target transverse  
spin would reveal unique information on the nature of the particles exchanged in the $u$ channel, be it 
a ``mere'' baryon slightly off-shell, or three perturbative quarks. For non-vanishing transverse momenta ($\Delta_T$),
 one expects in the latter case an asymmetry of the same order as the unpolarised cross section, while, in the former
case, they would be most likely decreasing for increasing $W^2$ and $Q^2$.

In conclusion, let us stress that  we expect spin observables to be a major tool to establish the dominance 
of a partonic process in the hard-exclusive reactions on which we have focused here. This reminds us of 
their importance in opening the field of Generalised Parton Distribution studies. Scaling tests --
 although crucial for the proof of the adequacy of the theoretical description  -- are often 
experimentally  undecisive because of the small lever arm reachable in pionneering experiments. Even 
in the case of unpolarised beam and targets, the dominance of transversely-polarised virtual-photon 
exchange may be checked  through azymuthal dependence studies.  As emphasised here, using
 transversely-polarised targets or analysing the polarisation of final-states particles will 
allow for other likely decisive tests of the QCD understanding of these reactions.

\section*{Acknowledgments}

We thank S.J. Brodsky, 
G. Huber, V. Kubarovsky, K.J. Park, B.~Pasquini, K. Semenov-Tian-Shansky, P. Stoler for useful and motivating discussions.
This work is partly supported by the ANR contract BLAN07-2-191986. L.Sz. acknowledges the support 
by the Polish Grant N202 249235.

\end{document}